\documentclass[12pt]{article}
\usepackage{epsfig}

\def\beq{\begin{equation}}
\def\eeq#1{\label{#1}\end{equation}}
\def\eeqn{\end{equation}}

\def\beqa{\begin{eqnarray}}
\def\eeqa#1{\label{#1}\end{eqnarray}}
\def\eeqan{\end{eqnarray}}

\let\bar=\overbar

\def\Dslash{\not{\hbox{\kern-4pt $D$}}}
\def\dslash{\not{\hbox{\kern-2pt $\del$}}}

\def\msb{{\bar{\ssstyle M \kern -1pt S}}}

\def\Kpnn {K^+ \to \pi^+ \nu \overline{\nu}}

\def\Bbar    {\kern 0.18em\overline{\kern -0.18em B}{}}
\def\Kbar    {\kern 0.18em\overline{\kern -0.18em K}{}}

\def\Title#1{\begin{center} { {\bf #1} } \end{center}}

\begin{document}

\Title{\Large ORKA: A Precision Measurement of 
\mbox{\boldmath$K^+ \to \pi^+ \nu \overline{\nu}$} at Fermilab}

\bigskip\bigskip


\begin{raggedright}  

{\it Jack L. Ritchie\index{Reggiano, D.}\\
Department of Physics \\
University of Texas at Austin\\
Austin, TX  78712, USA \\
Representing the ORKA Collaboration}
\bigskip\bigskip
\end{raggedright}

\noindent {Proceedings of CKM 2012, the 7th International Workshop on the CKM Unitarity 
Triangle, University of Cincinnati, USA, 28 September - 2 October 2012}

\section{Introduction}

\begin{figure}[htb]
\begin{center}
\epsfig{file=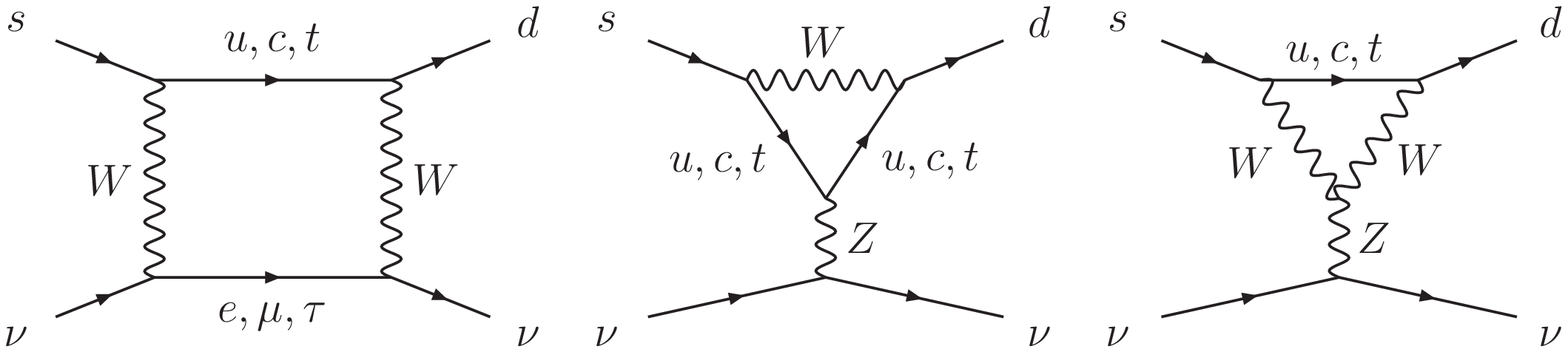,height=1.2in}
\caption{Electroweak loop diagrams responsible for $\Kpnn$.}
\label{fig:kpnndiags}
\end{center}
\end{figure}

\begin{figure}[htb]
\begin{center}
\epsfig{file=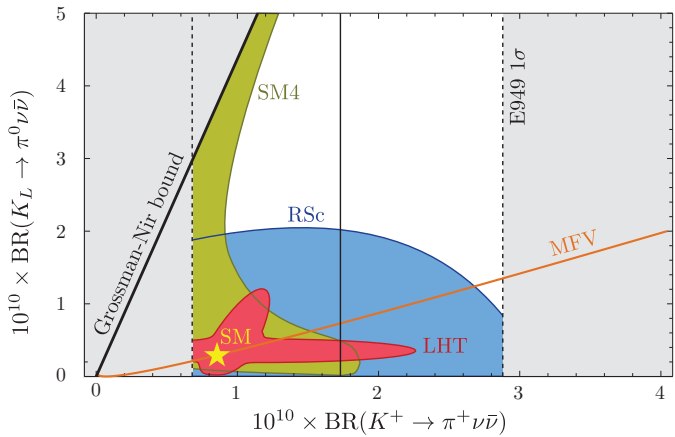,height=2.25in}
\caption{Predictions from various models for the branching fractions of $\Kpnn$
and $K_L^0 \to \pi^0 \nu \overline{\nu}$.  The yellow star shows the SM predictions.
The red  region is allowed by Littlest Higgs models with T-violation (LHT).
The blue region is  preferred by the Randall-Sundrum model with custodial protection (RSc).
The olive region is preferred by the Standard Model with a fourth
sequential generation (SM4).  The orange line shows the constraint imposed by minimal flavor
violation (MFV).  The vertical dashed lines show the $\pm 1 \sigma$ region from the
BNL E787/E949 $\Kpnn$ measurement.  The figure is 
reproduced from Reference~\ref{fig:straub}.
}
\label{fig:straub}
\end{center}
\end{figure}

The decay $\Kpnn$ is highly suppressed in the Standard Model (SM) and occurs through
the diagrams in Figure~\ref{fig:kpnndiags}.  
Several factors make this decay 
``theoretically clean."  In the usual operator product expansion treatment, a single
effective operator is relevant.  Loops involving the top quark dominate.  The hadronic
current is known from the well-measured $K_L^0 \to \pi^+ e^- \overline{\nu}$ decay. 
In fact, the dominant uncertainties on the $\Kpnn$ branching fraction in the SM
result from uncertainties on CKM elements, which
will be reduced over time.  The current SM expectation \cite{Brod} is
$$ {\cal B}(\Kpnn) = (7.8 \pm 0.8) \times 10^{-11}.
$$
Significantly, this decay remains clean in most extensions of the SM, and many new
physics scenarios can lead to deviations from the SM expectation by at least a factor of two.
Figure~\ref{fig:straub} shows expected branching fractions  in a few new
physics scenarios.

The $\Kpnn$ decay was observed by the Brookhaven experiments E787 and E949, which ran at the
Alternating Gradient Synchrotron.  E949 was an upgrade to the earlier E787, using the same
separated $K^+$ beamline and much of the same detector. The experiment used the stopped-$K^+$ 
technique.  E787/E949 observed \cite{E949} seven signal
events 
and reported a branching 
fraction measurement:
$$ {\cal B}(\Kpnn) = 1.73^{+1.15}_{-1.05}  \times 10^{-10}.
$$

A new experiment is being prepared at CERN, NA62, which will take data in the next few years in
parallel with LHC running.  NA62 builds upon a well-established kaon physics program (NA31, NA48)
and expects to be able to collect about 100 $\Kpnn$ events. NA62 will be the first experiment
to apply the decay-in-flight approach to the $\Kpnn$ decay.
NA62 is described in another contribution to CKM2012.

The ORKA experiment has been proposed \cite{ORKAproposal}
at Fermilab to extend the $\Kpnn$ reach to the 1000 event level.
This will provide a powerful test for new physics.

\section{The ORKA Experiment}

The basic concept of ORKA is to apply the method and techniques that were demonstrated
in BNL E787/E949.  ORKA does not  require better background rejection than E949
achieved.  ORKA will use existing facilities at Fermilab, including detector infrastructure
and a superconducting solenoid (from CDF) to moderate costs.  ORKA will be a fully
state-of-the-art detector, which will lead to substantial gains compared to E949 in 
performance and data-acquisition capability.  

\begin{figure}[htb]
\begin{center}
\epsfig{file=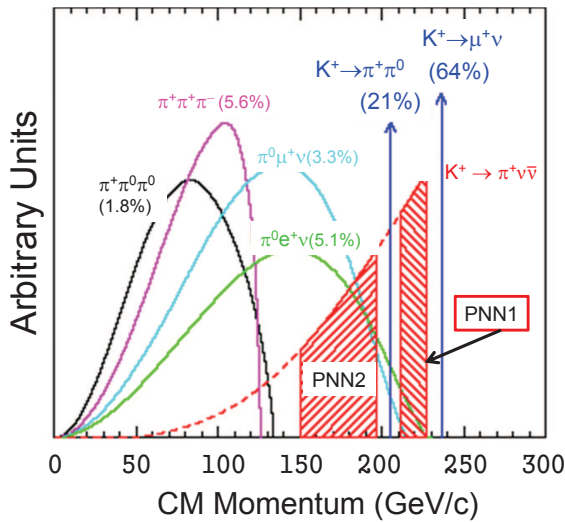,height=2.75in}
\caption{Charged particle momentum for the signal and
background decays.
Values in parentheses represent
the branching fractions of the decay modes. The regions indicated as PNN1 and PNN2
are signal regions used in the experiment.}
\label{fig:backgrounds}
\end{center}
\end{figure}

The signature for $\Kpnn$ is a single $\pi^+$ with no associated particles.  The major backgrounds
are the decays $K^+ \to \pi^+ \pi^0$, where both $\gamma$'s from the $\pi^0$ are missed, and
$K^+ \to \mu^+ \nu$, where the muon is misidentified as a pion.  The backgrounds are suppressed
by a hermetic photon veto system and extremely strong $\pi^+$ 
identification from observing the full decay chain $\pi \to \mu \to e$.  Figure~\ref{fig:backgrounds}
shows the $\pi^+$ (or $\mu^+$) momentum in the $K^+$ rest frame, which is the lab frame of the
experiment since beam $K^+$'s are stopped at the center of the detector in an active scintillating
fiber stopping-target.  
The experiment is sensitive to $\Kpnn$ events in two signal regions, PNN1
between the  $K^+ \to \pi^+ \pi^0$ and
$K^+ \to \mu^+ \nu$ peaks seen in Figure~\ref{fig:backgrounds}, and PNN2 below the $K^+ \to \pi^+ \pi^0$
peak.  With a high-statistics event sample, ORKA will measure not only the branching fraction but will also be sensitive to the effects of possible scalar or tensor interactions on the $\pi^+$ spectrum.

The $K^+$ beam for ORKA will be produced by $95 \, {\rm GeV}$ protons from the Fermilab Main Injector,
extracted over a $4.4 \, {\rm s}$ spill every $10 \, {\rm s}$, providing a beam power of about $75 \, {\rm kW}$. This is an increase of almost a factor of two over the beam power for E949, and it provides
about $9 \times 10^7$ $K^+$/spill into the acceptance of the secondary beamline.  The separated 
secondary beam will deliver a $K^+/\pi^+$ ratio of about three.  By running at lower momentum than the 
BNL experiment ($600 \, {\rm MeV}$ vs $710 \, {\rm MeV}$), the fraction of the $K^+$'s stopped 
in the active stopping-target
will more than double.  
When all anticipated beam changes are accounted for, ORKA will sample about a factor
of seven more stopped-$K^+$ decays per second than E949.
 
\begin{figure}[htb]
\begin{center}
\epsfig{file=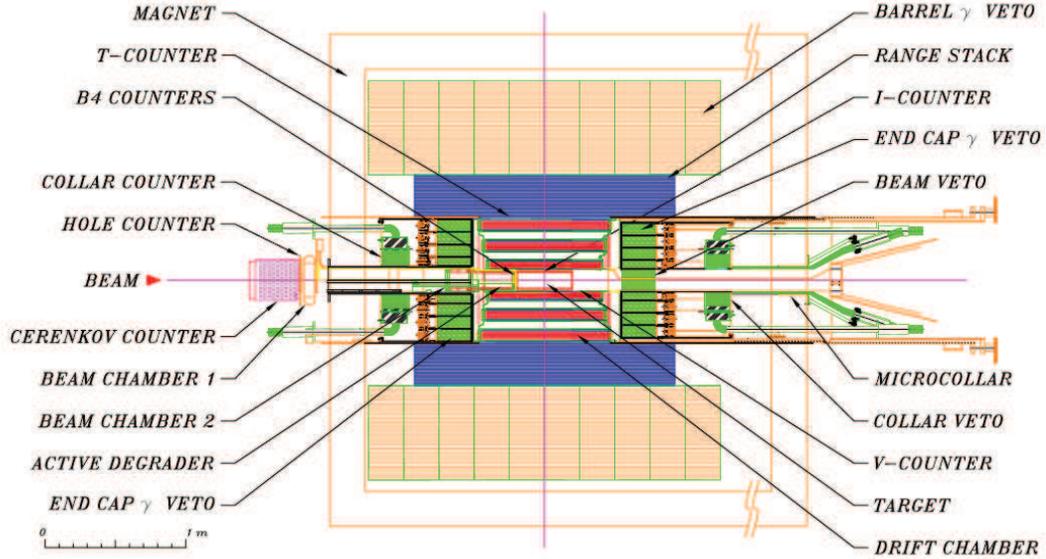,height=3.0in}
\caption{Elevation view of the proposed ORKA detector. Beam enters from the left.}
\label{fig:detector}
\end{center}
\end{figure}

The ORKA detector is shown in Figure~\ref{fig:detector}.  While modeled after E787/E949, all detector 
components will be new.  Large performance gains will come from several sources.  
A larger magnetic field 
($1.0 \, {\rm T} \to 1.25 \, {\rm T}$) will improve momentum resolution.  A longer drift chamber will increase geometrical acceptance.  An improved scintillating fiber stopping-target with smaller fibers
and better light collection will improve $K$:$\pi$ separation and improve vertex resolution.  A thicker
($17 X_0 \to 23 X_0$) photon veto system will improve $\pi^0$ rejection.
Waveform digitizers ($500 \, {\rm MHz}$, 10-bit) on all scintillators without multiplexing will reduce 
sensitivity to accidentals.  A modern ``triggerless" data-acquisition system will eliminate deadtime.
Extrapolations from E949 experience show a net gain in acceptance of a factor of about 11 over
E949, while maintaining as good or better background rejection.

\begin{figure}[htb]
\begin{center}
\epsfig{file=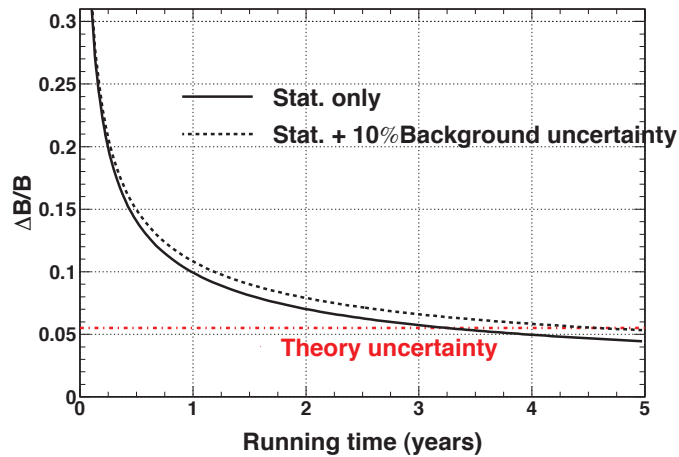,height=2.5in}
\caption{Fractional $\Kpnn$ branching fraction uncertainty versus
running time. Solid (dashed) curve is the sensitivity for no (10\%) background uncertainty.
The red (dot-dashed) line is the theory uncertainty excluding uncertainties on CKM elements.}
\label{fig:sensitivity}
\end{center}
\end{figure}

In 5000 hours of running per year, ORKA will collect about 210 $\Kpnn$ events per year if
the branching fraction is at the SM level.  The expected branching fraction uncertainty versus
time is shown in Figure~\ref{fig:sensitivity} assuming the
same signal/background ratio achieved in E949.  The experimental uncertainty
reaches the expected theory uncertainty after a few years of running.  In addition, 
ORKA will be able to make improvements in the measurement of several other $K^+$
decay modes.

\section{ORKA Status}

ORKA received Stage-1 approval from Fermilab in December, 2011,
and is conducting in R\&D in advance of DOE approval.   
A schedule with data taking early in 2017 is technically possible,
but is contingent on funding.
The ORKA Collaboration\footnote{The ORKA Collaboration consists of groups from
Arizona State University, University of British Columbia, Brookhaven National Lab, Fermilab, University of Illinois at Urbana-Champaign, INFN-Pisa, INFN-Napoli, INR-Moscow, JINR (Russia),
University of Mississippi, Notre Dame, University of Northern British Columbia, 
Universidad Nacional Autonoma de Mexico,  
Universidad Autonoma de San Luis Potosi (Mexico), 
University of Texas at Arlington, University of Texas at Austin,
TRIUMF, and Tsinghua University (China).
}
is very motivated, active, and hopeful that this experiment will proceed in a timely way.



%
%
%
%
 

\begin{thebibliography}{99}


\bibitem{Brod}
J. Brod, M. Gorbahn, and E. Stamou, Phys.\ Rev.\ D {\bf 83}, 034030 (2011).

\bibitem{Straub}
D. M. Straub, ``New physics correlations in rare decays," talk presented CKM2010, Warwick, UK, (September 6-10, 2010),
arXiv:1012.3893 [hep-ph].

\bibitem{E949}
A.V. Artamonov {\it et al.}, Phys.\ Rev.\ D {\bf 79}, 092004 (2009).

\bibitem{ORKAproposal}
http://projects-docdb.fnal.gov/cgi-bin/ShowDocument?docid=1365



\end{thebibliography}
\end{document}